\documentclass{pi2005}
\usepackage{amsmath,amsthm,amsfonts,amssymb,cite}
\usepackage{epsfig}

\slacs{.4ex}

\def\eqa{&=&}
\def\ccr{\nonumber\\}
\newcommand{\la}{\langle}
\newcommand{\ra}{\rangle}

\def\del{\Delta}
\def\ddel{{}^\bullet\! \Delta}
\def\deld{\Delta^{\hskip -.5mm \bullet}}
\def\dddel{{}^{\bullet \bullet} \! \Delta}
\def\ddeld{{}^{\bullet}\! \Delta^{\hskip -.5mm \bullet}}
\def\deldd{\Delta^{\hskip -.5mm \bullet \bullet}}
\def\bchi{\bar{\chi}}
\def\bpsi{\bar{\psi}}
\def\bep{\bar{\epsilon}}
\def\ep{\epsilon}
\def\Tr{\mathop{\mathrm{Tr}}\nolimits}
\def\Det{\mathop{\mathrm{Det}}\nolimits}

\begin{document}
\title{Path integrals in curved space and the worldline formalism}
\authori{Fiorenzo Bastianelli}
\addressi{Dipartimento  di Fisica, Universit{\`a} di Bologna
and INFN, Sezione di Bologna,\\ Via Irnerio 46, I-40126 Bologna, Italy}
\authorii{}     \addressii{}
\headauthor{Fiorenzo Bastianelli}
\headtitle{Path integrals in curved space and the worldline
formalism}
\lastevenhead{F. Bastianelli: Path integrals in curved space and
the worldline formalism}
 \maketitle

\begin{abstract}

We describe, how to construct and compute unambiguously path
integrals for particles moving in a curved space, and how these
path integrals can be used to calculate Feynman graphs and
effective actions for various quantum field theories with external
gravity in the framework of the worldline formalism. In
particular, we review a recent application of this worldline
approach and discuss vector and  antisymmetric tensor fields
coupled to gravity. This requires the construction of a path
integral for the $N=2$ spinning particle, which is used to compute
the first three Seeley--DeWitt coefficients for all $p$-form gauge
fields in all dimensions and to derive exact duality relations.
\end{abstract}

\section{Introduction}

The worldline formalism is an approach based on first quantization
which allows to obtain certain QFT results (amplitudes, effective
actions, etc.) in a rather simple way. It had been introduced by
Feynman in \cite{Feynman:1950ir} as ``an alternative to the
formulation of second quantization'', and by  Schwinger in his
famous paper on vacuum polarization \cite{Schwinger:1951nm}.
Feynman directly used a path integral approach to describe scalar
QED using worldlines of scalar particles, while Schwinger used
operatorial quantum mechanical methods to study vacuum effects in
QED. Also, the worldline formalism has served as a guide for
developing the first quantization of strings, from which it can be
recovered as point particle limit, hence the occasional name of
``string  inspired Feynman rules''. Many applications of this
formalism and references can be found in the review article
\cite{Schubert:2001he}. More recent applications include the
coupling to external gravitational fields
\cite{Ba:2002fv,Ba:2002qw,Ba:2004zp,Ba:2005vk}, studies of string
dualities \cite{Gopakumar:2003ns}, as well as numerical
simulations to address nonperturbative issues \cite{Gies:2005sb}.

In this talk we review the use of the worldline approach to
quantum field theories coupled to external gravity, and discuss
the main technical tool that is used in such an approach: the path
integral for a particle moving in a curved space. This is a
subject which has had a longwinded history, with some old
controversies fully resolved by now. We end with a brief
description of the worldline approach to vector and antisymmetric
tensor fields coupled to gravity, which exemplifies the
effectiveness of such an approach.

\section{The case of a scalar field coupled to gravity}

The simplest way to introduce the worldline formalism with
background gravity is to consider the example of a scalar field
$\phi$ coupled to the metric $g_{\mu\nu}$. The euclidean QFT
action reads
\begin{equation}
S[\phi,g]=\int\D^Dx\sqrt{g}\,\frac12\left(g^{\mu\nu}\partial_\mu\phi
\partial_\nu\phi+m^2\phi^2+\xi R\phi^2\right),
\label{uno}
\end{equation}
where $m$ is the mass of the scalar particle, $g_{\mu\nu}$ is the
background metric, and $\xi$ is a nonminimal coupling to the
scalar curvature $R$. The euclidean one-loop effective action
$\Gamma[g]$ describes all possible one-loop graphs with the scalar
field in the loop and any number of gravitons on the external
legs, see Fig. 1.

\begin{figure}[t]
\begin{center}
\scalebox{.8}{
\includegraphics*[122pt,632pt][208pt,720pt]{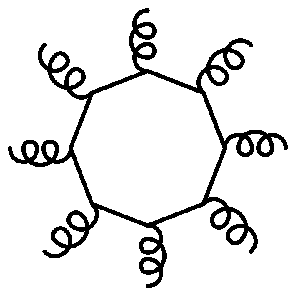}}
\end{center}
\vspace{-10mm} \caption{Loop of a scalar field with external
gravitons.}
\end{figure}

It can be obtained by path integrating the QFT action $S[\phi,g]$
over $\phi$, and  is formally given by
$$
\E^{-\Gamma[g]} \equiv \int {\cal D}\phi\,\E^{-S[\phi,g]}=
\Det^{-1/2} \left(-\nabla^2+m^2+\xi R\right),
$$
so that
\begin{eqnarray}
\Gamma[g] \eqa -\log\Det^{-1/2} (-\nabla^2+m^2+\xi R)=\ccr
\eqa\frac12\,\Tr\log(-\nabla^2 +m^2+\xi R)=\ccr \eqa
-\frac12\int_0^\infty\frac{\D T}T\,\Tr\E^{-T(-\nabla^2 +m^2 +\xi
R)}=\ccr \eqa-\frac12\int_0^\infty \frac{\D T}T \int_{T^1}{\cal
D}x\,\E^{-S[x;g]}\,, \label{wl}
\end{eqnarray}
where
$$
S[x;g]=\int_0^T\D\tau \left(\sfrac14\,g_{\mu\nu}(x)\dot x^\mu\dot
x^\nu+m^2+\xi R(x)\right).
$$
In the above equalities
$\nabla^2=g^{\mu\nu}\nabla_\mu\partial_\nu$ is the covariant
laplacian acting on scalars. In the third line of (\ref{wl}) we
have used the proper time representation of the logarithm,
$$
\log\frac{a}b=-\int_0^\infty \frac{\D
T}T\left(\E^{-aT}-\E^{-bT}\right),
$$
and dropped an additive constant. This provides the starting point
of the heat kernel method, originally due to Schwinger and which
is by now a well-appreciated tool for studying QFT in curved
backgrounds \cite{DW,DeWitt:2003pm}. In this method the operator
$\hat H=-\nabla^2 +m^2 +\xi R$ is reinterpreted as the quantum
hamiltonian of a ``fictitious'' mechanical model: that of a
nonrelativistic particle in curved space with a specific coupling
to the scalar curvature. The corresponding Schr\"odinger equation
is then used in trying to solve the problem. However, it proves
quite useful to reformulate this quantum mechanics using a path
integral: this is shown by the last equality in (\ref{wl}). The
exponent of the path integral contains the classical action of the
mechanical model whose quantization is expected to produce the
quantum hamiltonian $\hat H$. The operatorial trace is obtained by
using periodic boundary conditions on the worldline time $\tau\in
[0,T]$, which therefore describes a one-dimensional torus, or
circle, $T^1$.

It is clear that to use this final path integral formulation
one has to be able to define and compute
path integrals for particles moving in curved spaces quite precisely.
This has been a notoriously complicated and controversial subject.
However, this topic is now mature and solid, and
will be reviewed in the next section.

For studying other QFT models it is useful to note that the
previous effective action for the scalar field can be obtained by
first quantizing a scalar point particle with coordinates $x^\mu$
and auxiliary einbein $e$, which is described by the action
\cite{Brink:1976sz}
$$
S[x^\mu,e]=\int_0^1\D\tau\,\frac12\,\Bigl[e^{-1} g_{\mu\nu}(x)
\dot x^\mu \dot x^\nu+e\bigl(m^2 + \xi R(x)\bigr)\Bigr].
$$
This worldline action is reparametrization invariant. One can
eliminate almost completely the einbein by the gauge condition
$e(\tau)=2T$, and integrate over the remaining modular parameter
$T$ after taking into account the correct measure. This reproduces
the previous answer in (\ref{wl}). Thus the ``fictitious'' quantum
mechanics mentioned above is not at all that fictitious: it
corresponds to the first quantization of the scalar particle which
makes the loop in the Feynman graph of Fig. 1. This picture can be
enlarged to include particles with spin. Extending the worldline
symmetry to $N=1$ supergravity gives a description of a spin
$\frac12$ particle, while $N=2$ supergravity describes particles
associated to vector and antisymmetric tensor fields
\cite{Brink:1976sz,Brink:1976uf,Berezin:1976eg,Gershun:1979fb,Howe:1988ft}.
A slightly different approach was discussed in \cite{Halpern:1977he}.
Gauge fixed versions of these particle models were in fact used to
compute gravitational and chiral anomalies in one of the most
beautiful applications of the worldline approach
\cite{Alvarez-Gaume:1983at,Alvarez-Gaume:1983ig,Friedan:1983xr}.

Other applications of this worldline approach with external
gravity include the computation of trace anomalies
\cite{Ba:1991be,Ba:1992ct,Ba:2000dw,Ba:2001tb}, which in fact was
one of the main motivations to study anew path integrals in curved
spaces, as well as the calculation of some amplitudes, like the
one-loop correction to the graviton propagator due to loops of
spin 0, $\frac12$, 1 and antisymmetric tensor fields
\cite{Ba:2002fv,Ba:2002qw,Ba:2005vk} (see Fig. 2).

\begin{figure}[t]
\begin{center}
\includegraphics*[100pt,667pt][241pt,725pt]{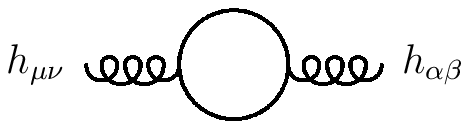}
\raisebox{-.5cm}{
\includegraphics*[100pt,652pt][221pt,725pt]{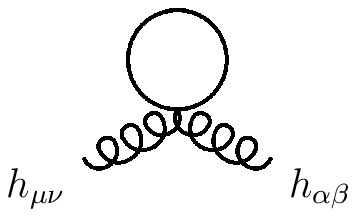}
}
\end{center}
\vspace{-10mm} \caption{One-loop matter contributions to the
graviton 2-point function.}
\end{figure}

In \cite{Ba:2004zp} the one-loop corrections to the
graviton-photon mixing in constant electromagnetic fields due to
virtual charged particles has been computed (see Fig. 3). This
calculation would have been very difficult to perform using
standard methods.

\begin{figure}[t]
\begin{center}
\includegraphics*[100pt,667pt][241pt,725pt]{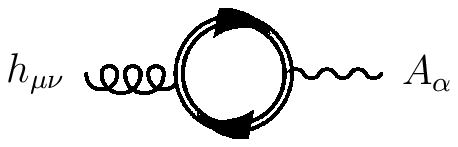}
\end{center}
\vspace{-10mm} \caption{One-loop correction to graviton-photon
mixing in a constant electromagnetic field.}
\end{figure}

\section{Path integrals in curved space: regularizations and counterterms}

In this section we wish to discuss, how to construct and compute
path integrals for a nonrelativistic particle moving in a curved
space. With a slight change in notations (we consider a particle
of unit mass which propagates for a total time $\beta$, so to have
a standard  normalization of the action) we consider the following
euclidean action
\begin{equation}
S[x]= \int_0^{\beta}\D t\left(\sfrac12\,g_{\mu\nu}(x)\dot x^\mu
\dot x^\nu +V(x)\right),
\label{nlsm}
\end{equation}
where $V$ is an arbitrary scalar potential, from which one wants
to construct the path integral
\begin{equation}
Z=\int{\cal D}x\,\E^{- S[x]}\,.
\label{aim}
\end{equation}
Construction and computations of path integrals for these
nonlinear sigma models can be quite subtle. For example one may
find the following description in a well-known textbook on path
integrals \cite{LS}: ``\textit{If you like excitement, conflict
and controversy \ldots{} then you will love the history of
quantization on curved spaces. \ldots{} people continue to get
signs and factors of 2 wrong in their results.}'' That was surely
a fair description of the situation at the time that book was
written, but now all major difficulties are understood and taken
care of, as we are going to describe next.

One dimensional nonlinear sigma models suffer from ordering
ambiguities when one applies canonical quantization. The classical
hamiltonian reads
$$
H=\sfrac12\,g^{\mu\nu}(x) p_\mu p_\nu +V(x)
$$
and one has to specify an ordering between the $p$'s and the
$x$'s. One can impose covariance under change of coordinates at
the quantum level, but this selects a subclass of  possible
orderings and leave unfixed an arbitrary coupling to the scalar
curvature $R$ (the only scalar object that one can construct with
two derivative on the metric is the scalar curvature). Thus in the
coordinate representation ($p_\mu\to-\I\partial_\mu$) one has a
family of covariant quantum hamiltonians
$$
\hat H=-\sfrac12\,\nabla^2+\alpha R+V(x)\,,
$$
which depend on the parameter $\alpha$. In the absence of other
symmetries that can be used to identify a unique quantum theory,
one has to extract the value of $\alpha$ from ``experiments'',
i.e. from the particular physical problem one wishes to describe
with the sigma model. For example, in the case discussed in
section 2, the scalar relativistic particle, one may demand that
conformal invariance holds for vanishing mass, thus fixing
$\alpha=\xi/2=(D-2)/8(D-1)$ 
in $D$ dimensions.
Given this situation, one can always decide to set $\alpha=0$, and
describe additional couplings to $R$ as extra terms contained in
the potential $V$. This is what we will do in the following. The
requirement that $\hat H$ be covariant and without any coupling to
$R$ will be the ``renormalization conditions'' which will be
imposed on the path integral.
Equivalently, these conditions can be imposed at the level of the 
effective action
$$
\Gamma = -\log Z = \cdots - \frac{\beta}{12} R + \cdots  
$$
where $Z$ is the path integral given in (\ref{aim}).

This path integral can be dealt with just as higher dimensional
path integrals, i.e. QFT path integrals, where renormalization is
needed. Indeed one can always imagine quantum mechanics as a $0+1$
dimensional QFT. Then to compute these path integrals one must use
a regularization scheme which consist of: i) a
\underline{regularization}, ii) suitable
\underline{renormalization conditions}, and iii) \underline{local
counterterms} needed to satisfy the renormalization conditions and
to eliminate any source of ambiguity. This way any regularization
scheme will produce the same correct final answer.

To recognize why a renormalization scheme is necessary, it is
enough to notice that the nonlinear sigma model in (\ref{nlsm})
has derivative interactions which seem to give rise to linear
divergences. These divergences can be renormalized away, but that
is not necessary. In fact, the covariant path integral measure
produces local interactions with additional linear divergences
that cancel the previous ones \cite{Lee:1962vm}. Thus,
one-dimensional nonlinear sigma models are finite. Nevertheless
one needs a regularization scheme  to handle intermediate
divergences and ambiguities. Counterterms are then used to satisfy
the renormalization conditions. Since after all the theory is
finite, these counterterms are also finite. (One may consider the
measure as giving for free the infinite part of the counterterms).
Power counting shows that  one-dimensional nonlinear sigma models
are super-renormalizable, and thus counterterms can only appear up
to two-loops.

The previous discussion can be readily exemplified. Let us Taylor
expand the metric around the origin
$g_{\mu\nu}(x)=g_{\mu\nu}(0)+x^\alpha\partial_\alpha
g_{\mu\nu}(0)+\ldots$ and insert this expansion into the action
(\ref{nlsm}). From the leading constant term $g_{\mu\nu}(0)$ one
obtains the propagator, which in momentum space and for large
momentum $k$ goes like
$$
\la x(k)x(-k) \ra \sim k^{-2}\,.
$$
Then the next term in the expansion of the metric gives a
trilinear vertex with two derivatives of the type $x\dot x^2$ (we
will indicate each derivative by a dot also in Feynman diagrams),
so that one can construct the following linearly divergent graph
(graphs on the worldline, i.e. in the $0+1$ dimensional QFT)
$$
\raisebox{-.7cm}{\scalebox{.8}{
{\includegraphics*[122pt,663pt][220pt,715pt]{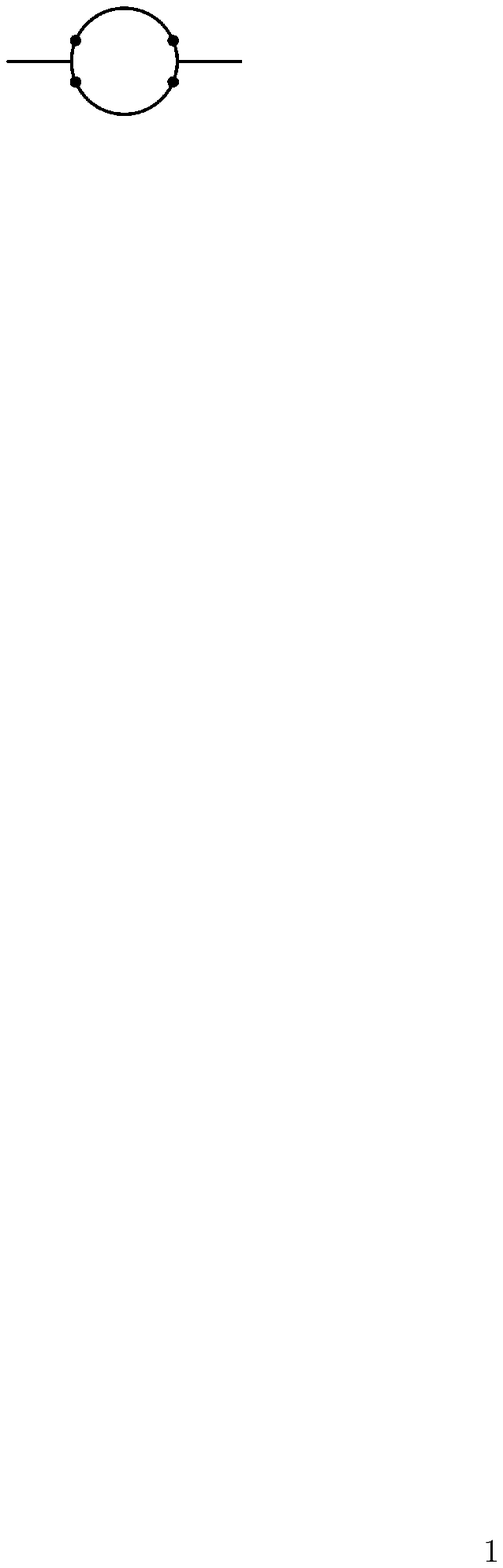}}}}
\ \sim \ \int\D k\ \frac{k^4}{k^4}\  \to \ \mbox{linear\
divergence.}
$$
The propagators are in fact compensated by the derivatives that
act on each vertex, and this gives rise to a linear divergence.

However, one should  consider that the covariant path integral
measure carries extra terms. The covariant measure in (\ref{aim})
is formally given by
$$
{\cal D}x \sim \prod_t \sqrt{g(x(t))}\,\D^D x(t)\,,
$$
but one can exponentiate the nontrivial $\sqrt{g}$ dependence
using commuting $a^\mu$ and anticommuting $b^\mu$, $c^\mu$ ghost
fields with action
$$
S_{\mathrm{measure}}[x,a,b,c]=\int_0^{\beta}\D t\,
\sfrac12\,g_{\mu\nu}(x)(a^\mu a^\nu+b^\mu c^\nu)\,,
$$
so that path integrating over these ghosts reproduces the correct
measure. The advantage of this exponentiation is that one can
consider perturbatively the effect of the measure, and recognize
to which type of diagram they contribute to. One can use the
leading term of the Taylor expansion of the metric in the ghost
action to identify the ghost propagators, which for large momenta
go like
$$
\la a(k) a(-k) \ra \sim \la b(k) c(-k) \ra \sim 1\,.
$$
The next term produces a vertex where $x$ couples to the ghosts,
and one obtains again a linearly divergent diagram of the type
$$
\raisebox{-.7cm}{\scalebox{.8}{
{\includegraphics*[122pt,663pt][220pt,715pt]{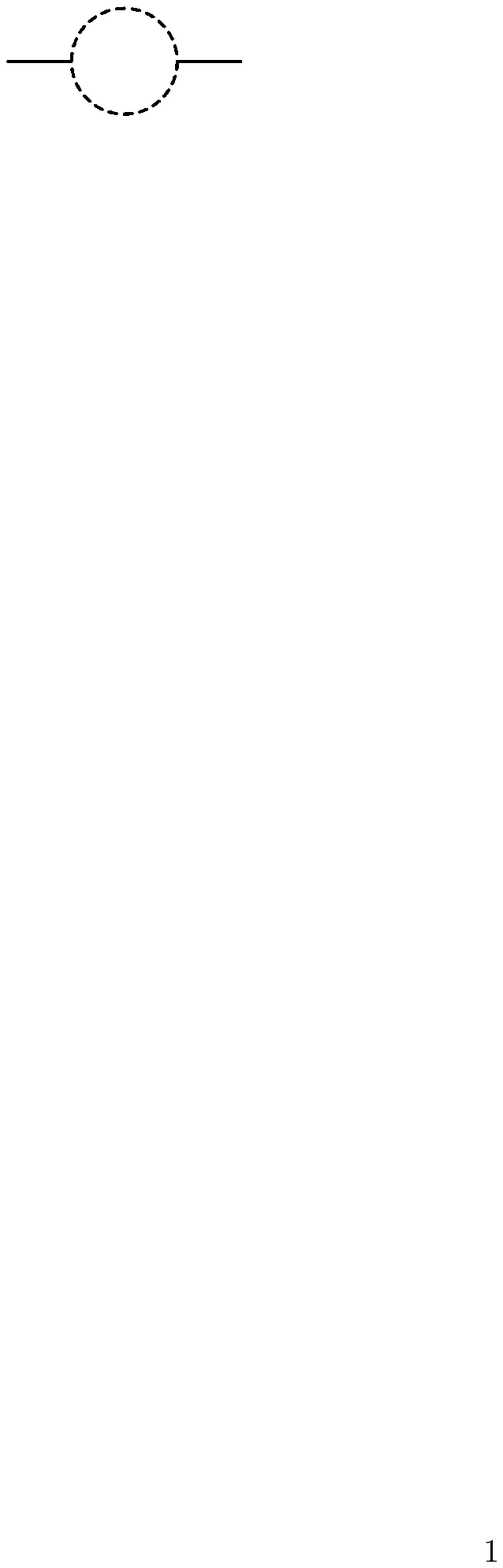}}}}
\ \sim \ \int\D k\  \to \ \mbox{linear\ divergence,}
$$
where dashed lines denote ghost propagators. One may check that
the previous two diagrams combine to produce a finite result
$$
\raisebox{-.7cm}{\scalebox{.8}{
{\includegraphics*[122pt,663pt][220pt,715pt]{bastianelli5.eps}}}}
+ \!\!\! \raisebox{-.7cm}{\scalebox{.8}{
{\includegraphics*[122pt,663pt][220pt,715pt]{bastianelli6.eps}}}}=\mbox{finite.}
$$
The cancellation must of course be achieved carefully: one must
regulate each diverging graph and then combine them. Only at this
stage one is allowed to remove the regulator. Different regulators
may lead to different left over finite parts. Then different
counterterms associated to different regularization schemes make
sure that this difference is accounted for to obtain the same
correct final answer.

Three different regularization schemes have been developed and
checked thoroughly: mode regularization (MR)
\cite{Ba:1991be,Ba:1998jm}, time slicing (TS)
\cite{DeBoer:1995hv}, and dimensional regularization (DR)
\cite{Ba:2000nm}. The precise details of each regularization
scheme  can be found in the literature, and we will give here only
a brief description for each of them.

Mode regularization starts by expanding all fields in Fourier
sums. The regularization is achieved by truncating these sums at a
fixed mode $M$, so that all distributions that appear in Feynman
graphs become well-behaved functions. Then one performs all
computations at finite $M$, as they are now completely unambiguous
(one may check for example that after including the ghosts all
possible divergences cancel). Eventually one takes the limit
$M\rightarrow \infty$, thus obtaining a unique finite result. In
practice one can proceed faster: one may perform all manipulations
that are valid at the regulated level (for example partial
integration) to cast the integrands in alternative forms that can
be computed directly in the $M \to \infty$ limit. This scheme
requires the addition of a local counterterm  $V_{\mathrm{MR}}$ to
the action (\ref{nlsm}) to satisfy the renormalization conditions
mentioned earlier. This local counterterm  is given by
$$
 V_{\mathrm{MR}}=-\frac18\,R-\frac1{24}\,g^{\mu\nu} g^{\alpha \beta}
g_{\gamma\delta}\,\Gamma_{\mu\alpha}^\gamma\Gamma_{\nu\beta}^\delta\,.
$$
The noncovariant piece is necessary to restore covariance (which
is broken at the regulated level), so that the complete final
result is covariant. This regularization is analogous to the
standard momentum cut-off used in quantum field theories.

Time slicing is a regularization that is derived from the exact
operatorial expression of the transition amplitude. Inserting
completeness relations and using the ``mid-point prescription''
(related to the Weyl ordering of the operators), one derives a
discretized path integral in momentum space. By integrating out
the momenta and taking the continuum limit, one carefully derives
the prescriptions needed for evaluating consistently the products
of distributions contained in Feynman diagrams
\cite{DeBoer:1995hv}. In particular, the Heaviside step function
acquires the value $\theta(0)=\frac12$, while Dirac deltas must be
used as Kronecker deltas. This regularization requires the
counterterm
$$
V_{\mathrm{TS}}=-\frac18\,R+\frac18\, g^{\mu\nu}\,
\Gamma_{\mu\alpha}^\beta \Gamma_{\nu\beta}^\alpha\,,
$$
which is seen to arise from Weyl ordering the quantum hamiltonian
\cite{Gervais:1976ws,Sakita:1986ad}. Time slicing is a
regularization that can be considered analogous to lattice
regularization of usual quantum field theories. In
\cite{Ba:1998jb} it was checked that MR and TS give the same
result for the transition amplitude to order $\beta^2$, where
$\beta$ is the total propagation time. That calculation produced
as byproduct the first three Seeley--DeWitt coefficients for a
scalar particle including the corrections for noncoinciding
points.

Dimensional regularization is a perturbative regularization which
uses an adaptation of standard dimensional regularization to
regulate the distributions defined on the compact space
$I=[0,\beta]$. One adds $d$ extra infinite dimensions
$I\rightarrow I\times R^d\equiv\Omega$ and perform all
computations of ambiguous Feynman graphs in $d+1$ dimensions.
Extra dimensions act as a regulator when $d$ is extended
analytically in the complex plane, as usual. After evaluation of
the integrals one should take the $d\rightarrow 0$ limit. In fact,
this is quite difficult since the compact space $I$ produces sums
over discrete momenta, and the standard formulas of dimensional
regularization do not include such a situation. However there is
no need to compute at arbitrary complex $d$. One may use
manipulations valid at the regulated level, like differential
equations satisfied by the Green functions and partial
integration, to cast the integrand in equivalent forms that, on
the other hand, can be unambiguously computed in the $d\to0$
limit. This method carries a covariant counterterm
\begin{equation}
V_{\mathrm{DR}}=-\frac18\,R\,.
\label{ctdr}
\end{equation}
For sigma models with infinite propagation time one can use the
standard formulas of dimensional regularization, and in
\cite{Kleinert:1999aq} it was originally understood that noncovariant
counterterms did not arise. In \cite{Ba:2000pt} it was checked
that this counterterm is covariant and given by (\ref{ctdr}).

The previous discussions can be further clarified by going through a
specific example. Consider the following
superficially logarithmic divergent graph $G$
$$
G = \raisebox{-.7cm}{\scalebox{.8}{
{\includegraphics*[147pt,663pt][197pt,715pt]{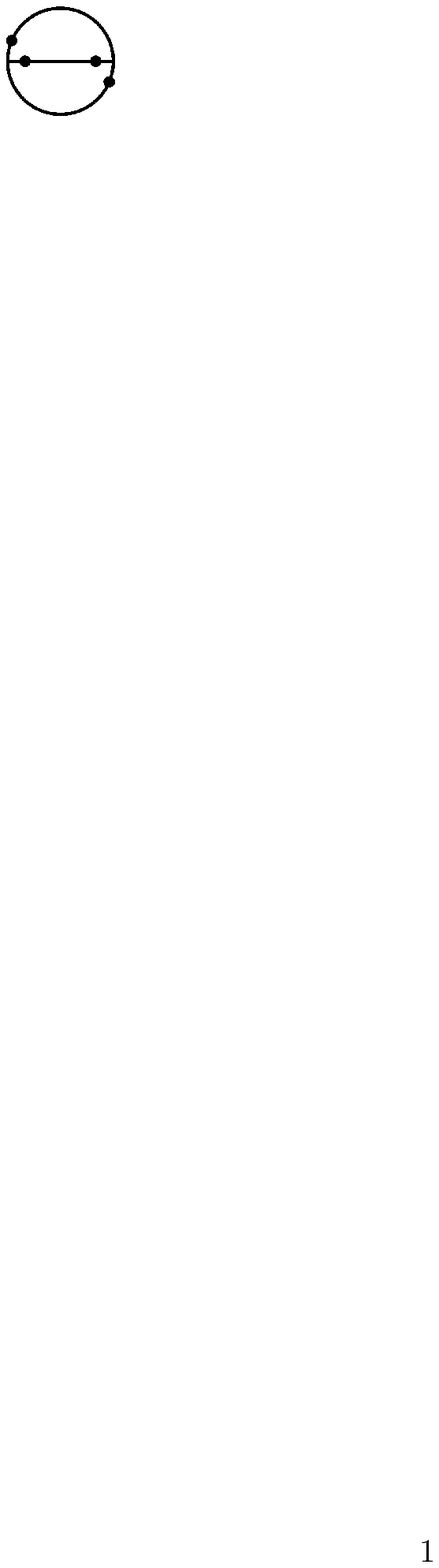}} }}
= \int_0^1\D\tau \! \int_0^1\D\sigma \ \ddel \ \ \ddeld \ \, \deld
\,.
$$
In this example we use Dirichlet boundary conditions $x(0)=x(1)=0$
for the field $x(\tau)$, where $\tau={t/\beta}\in [0,1]$, so that
the propagator reads
$$
\la x(\tau)x(\sigma)\ra=-\beta\,\Delta(\tau,\sigma)
$$
with
\begin{eqnarray*}
\Delta(\tau,\sigma)&=&\sum_{m=1}^{\infty}\biggl[-\frac2{\pi^2m^2}\,
\sin(\pi m \tau) \sin (\pi m \sigma)\biggr]=\\
&=&(\tau-1)\sigma\,
\theta(\tau-\sigma)+(\sigma-1)\tau \,\theta(\sigma-\tau)\,,\nonumber\\
\ddel(\tau,\sigma)&=&\sum_{m=1}^{\infty} \biggl[-\frac2{\pi m}\,
\cos(\pi m\tau) \sin(\pi m \sigma)\biggr]=
\sigma-\theta(\sigma-\tau)\,,\nonumber \\
\ddeld(\tau,\sigma)&=&\sum_{m=1}^{\infty}\Bigl[-2\cos(\pi m\tau)
\cos(\pi m\sigma) \Bigr] =  1-\delta(\tau-\sigma)\,,
\end{eqnarray*}
where dots on the left/right indicate derivatives with respect to
the first/second variable.

\smallskip
\noindent $\bullet$ In mode regularization one cuts off the mode
expansion at a big mode number $M$, and proceeds as follows
\begin{eqnarray*}
 G({\rm MR})&=&\int_0^1 \D\tau \int_0^1 \D\sigma\
\ddel \ \, \ddeld \  \deld= \frac12\int_0^1 \D\tau
\int_0^1\D\sigma \
 \partial_\sigma (\ddel^2) \, \deld=\\
&=&-\frac12\int_0^1\D\tau \int_0^1\D\sigma\ \ddel^2 \, (\deldd) =
-\frac12\int_0^1\D\tau\int_0^1\D\sigma \ \ddel^2 \, (\dddel)=\\
&=&-\frac12\int_0^1\D\tau\int_0^1\D\sigma \
\frac13\,\partial_\tau(\ddel^3)= -\frac16\int_0^1\D\sigma \,
[\ddel^3(1,\sigma)-\ddel^3(0,\sigma)]\to\\
&\to&-\frac16\int_0^1\D\sigma\, [(\sigma-1)^3-\sigma^3]=
-\frac1{12}\,.
\end{eqnarray*}

\noindent $\bullet$ In time slicing one can use
$\theta(0)=\frac12$, and thus
$$
 G(\mathrm{ TS})=\int_0^1\D\tau\int_0^1\D\sigma \
\Bigl(\sigma-\theta(\sigma-\tau)\Bigr)\Bigl(1-\delta(\tau-\sigma)\Bigr)
\Bigl(\tau-\theta(\tau-\sigma)\Bigr)=-\frac16\,.
$$

\noindent $\bullet$ In dimensional regularization one extends the
action to higher dimensions as
$$
S =\int_0^1\D\tau\,\sfrac12\,g_{\mu\nu}(x)\dot x^\mu\dot x^\nu+
\ldots\Rightarrow\int_{\Omega}\D^{d+1}t\,\sfrac12\,g_{\mu\nu}(x)
\partial_a x^\mu \partial_a x^\nu +\ldots\,,
$$
where the repeated index $a$ is summed from 1 to $d+1$. From this
extended action one obtains vertices and propagators, and thus
\begin{eqnarray*}
 G(\mathrm{DR})&=&
\int\D^{D+1}t\int\D^{D+1}s\,(_a{\del})~(_a{\del_b})~(\del_b)=\\
&=&\int\D^{D+1}t\int\D^{D+1}s\,(_a{\del})~\partial_a\left(\sfrac12\,(\del_b)^2\right)=\\
&=&-\frac12\int\D^{D+1}t\int\D^{d+1}s\,(_{aa}{\del})~(\del_b)^2=\\
&=&-\frac12\int\D^{D+1}t\int\D^{D+1}s\,\delta^{D+1}(t,s)~(\del_b)^2=\\
&=&-\frac12\int\D^{D+1}t\,(\del_b)^2|_t\;\;\to\;\;
-\frac12\int_0^1\D\tau\;(\deld)^2\Bigl|_\tau=-\frac1{24}\,,
\end{eqnarray*}
where the vertical bar indicates evaluation at coinciding points,
and $_{a}{\del}\equiv\partial_a{\del}$. In this computation we
used the Green equation $_{aa}{\del}(t,s)=\delta^{d+1}(t,s)$
satisfied by the propagator in $d+1$ dimensions.

We have seen concretely, how different regularizations produce
different answers. However the different counterterms make sure
that the final  complete result is independent of the
regularization chosen. The conclusion is that path integrals in
curved spaces can be defined and computed without any ambiguity.

Let us conclude with another quote, now from a recent book of
DeWitt \cite{DW2}, which comments on the extra $R$ terms that
appear in the action (the terms that we now call counterterms):
``\textit{Many years ago the author was guilty of suggesting that
this term is $\frac16\,R$, a suggestion that remained in the
literature for a long time. That the term must be $\frac18\,R$ is
conclusively demonstrated in reference \cite{DeWitt:1992cy} where
the path integral derivation of the Chern--Gauss--Bonnet formula
for the Euler--Poincar\`e characteristic demands for its
consistency.}'' This statement witnesses the long lasting
confusion on how to calculate in a correct way path integrals  in
curved spaces. At the same time this statement is rather
misleading, as it  does not specify  how the path integral is
computed, i.e. which regularization scheme is used. Most likely
DeWitt had in mind a kind of covariant regularization similar to
DR (the difference in sign is due to different conventions adopted
in the definition of the curvature scalar).

Extensive descriptions, tests and applications of the previous
regularization schemes can be found in a forthcoming book
\cite{book}.

\section{\bmth{N=2} spinning particles and antisymmetric tensor fields }

We now describe a recent application of path integrals in curved
spaces: the worldline approach to vector and antisymmetric tensor
fields coupled to gravity \cite{Ba:2005vk}. The worldline action
that describes these models is given by the $N=2$ spinning
particle with quantized Chern--Simons coupling
\cite{Brink:1976uf,Berezin:1976eg,Gershun:1979fb,Howe:1988ft}.
This particle is described by phase space coordinates
$X=(x^\mu,p_\mu,\psi^\mu,\bar\psi_\mu)$ and gauge fields
$G=(e,\chi,\bar\chi,a)$. The variables $ \psi^\mu$,
$\bar\psi_\mu$, $\chi$, $\bar\chi$ are Grassmann variables. The
worldline action in flat $D$ dimensional target space is given by
$$
S[X,G]=\int\D t\Bigl(p_\mu\dot x^\mu+\I\bpsi{}_\mu\dot \psi^\mu-
eH-\I\bchi Q-\I\chi\bar Q-a(J-q)\Bigr),
$$
where the $N=2$ supersymmetry generators
$$
H=\sfrac12\,p_\mu p^\mu\,,\quad Q=p_\mu\psi^\mu\,,\quad
\bar{Q}=p_\mu\bar\psi^\mu\,,\quad J=\bar\psi^\mu\psi_\mu
$$
satisfy a first class Poisson-bracket algebra
$$
\{Q,\bar Q\}_{\mathrm{PB}}=-2\I H\,,\quad \{J,Q\}_{\mathrm{PB}}=\I
Q \,,\quad \{J,\bar Q\}_{\mathrm{PB}}=-\I\bar Q
$$
and are gauged by the Lagrange multipliers $G$. The Chern--Simons
coupling $q$ is quantized as $q=\frac12 D-p-1$, with $p$ an
integer. This model describes a $p$-form gauge field $A_p$ with
field strength $F_{p+1}=\D A_p$ and standard Maxwell action
\begin{equation}
S^{\mathrm{QFT}}_p=\int\D^Dx\,\frac1{2(p+1)!}\,F_{\mu_1\cdots\mu_{p+1}}F^{\mu_1\cdots\mu_{p+1}}\,.
\label{pact}
\end{equation}
This is immediately seen in canonical quantization, which is
introduced by interpreting the phase space coordinates as
operators with (anti)commutation relations
$$
[\hat x^\mu,\hat p_\nu]=\I\delta^\mu_\nu\,,\quad
\{\hat\psi^\mu,\hat\psi^\dagger_\nu\}=\delta^\mu_\nu\,.
$$
These operators act on wave functions $\phi$ which depend on the
classical configuration space coordinates $x^\mu$ and $\psi^\mu$,
and thus have an expansion of the form
$$
\phi(x,\psi)=F(x)+F_\mu(x)\psi^\mu+\frac12\,F_{\mu_1\mu_2}(x)\psi^{\mu_1}\psi^{\mu_2}+
\ldots+\frac1{D!}\,F_{\mu_1\ldots\mu_D}(x)\psi^{\mu_1}\cdots\psi^{\mu_D}\,.
$$
The first class constraints now become differential operators
$$
\begin{array}{rclrcl}
\hat H&=&-\dfrac12\,\partial_\mu\partial^\mu\,,\qquad&
\hat{Q}&=&-\I\psi^\mu\partial_\mu\,,\\[6pt]
\hat{Q}^\dagger&=&-\I\partial_\mu\,\dfrac{\partial}{\partial\psi_\mu}\,,\qquad&
\hat{J}&=&-\dfrac12\biggl[\psi^\mu,\dfrac{\partial}{\partial\psi^\mu}\biggr],
\end{array}
$$
which select the physical sector of the Hilbert space
\begin{eqnarray*}
(\hat J-q)\phi_{\mathrm{phys}}=0&\;\Rightarrow\;&
\phi_{\mathrm{phys}}\sim F_{\mu_1\ldots\mu_{p+1}}(x)
\psi^{\mu_1}\ldots\psi^{\mu_{p+1}}\,,\\
\hat Q\phi_{\mathrm{phys}}=0&\;\Rightarrow\;& \D F_{p+1}=0\,,\\
\hat{\bar Q}\phi_{\mathrm{phys}}=0&\;\Rightarrow\;& \D^\dagger
F_{p+1}=0\,.
\end{eqnarray*}
Thus ones sees that only the tensor $F_{\mu_1\ldots\mu_{p+1}}$
with $p+1$ indices is physical, and must satisfy the Bianchi
identities and Maxwell equations. Thus one concludes that the
physical sector of the $N=2$ spinning particle describes the field
strength of a $p$-form gauge field with the standard Maxwell
action (\ref{pact}).

To introduce the coupling to gravity, we couple the spinning
particle to a target space metric $g_{\mu\nu}$ (and corresponding
vielbein $e_\mu^a$)  preserving the $N=2$ local supersymmetry,
then go to configuration space by eliminating $p_\mu$, Wick rotate
to euclidean time ($t\to-\I\tau$, and also $a\to\I a$), and obtain
the euclidean action
\begin{eqnarray*}
S[X,G;g_{\mu\nu}]&=&\int_0^1\D\tau\biggl[\frac12\,e^{-1}g_{\mu\nu}
\left(\dot x^\mu-\bchi\psi^\mu-\chi\bpsi^\mu\right)
\left(\dot{x}^\nu-\bchi\psi^\nu-\chi\bpsi^\nu\right)+\\
&&+\bpsi_a\left(\dot\psi^a+\dot x^\mu\omega_\mu{}^a{}_b\psi^b+ \I
a\psi^a\right)-\frac{e}2\,R_{abcd}\bar\psi^a\psi^b
bar\psi^c\psi^d-\I qa\biggr].
\end{eqnarray*}
The gauge symmetries on the gauge multiplet $G$ are given by
\begin{equation}
\begin{array}{rcl}
\delta e&=&\dot\xi+2\bchi\ep+2\chi\bep\,,\\[4pt]
\delta\chi&=&\dot\ep+\I a\epsilon-\I\alpha\chi\,,\\[4pt]
\delta\bchi&=&\dot{\bep}-\I a\bep+\I\alpha\bchi\,,\\[4pt]
\delta a&=&\dot\alpha
\end{array}
\label{33}
\end{equation}
and do not couple to the target space geometry.

The one-loop effective action for a $p$-form gauge potential has
then the following worldline representation
$$
\Gamma^{\mathrm{QFT}}_{p}[g_{\mu\nu}]\sim  \int_{T^1}
\frac{\mathcal{D}G\,\mathcal{D}X}{\mathrm{Vol(Gauge)}}\,\E^{-S[X,G;g_{\mu\nu}]},
$$
but first one should fix the gauge symmetries (\ref{33}). On the
one-dimensional torus $T^1$ we adopt antiperiodic boundary
conditions for all fermionic fields. We choose the gauge
$\hat{G}=(\beta,0,0,\phi)$, insert the Faddeev--Popov
determinants, and integrate over the remaining moduli $\beta$ and
$\phi$. Fixing appropriately the overall normalization gives
\begin{equation}
\Gamma^{\mathrm{QFT}}_{p}[g_{\mu\nu}]=
-\frac12\int_0^\infty\frac{\D\beta}{\beta}
\int_0^{2\pi}\frac{\D\phi}{2\pi}\biggl(2\cos\frac{\phi}2\biggr)^{-2}
\int_{T^1}{\cal D}X\,\E^{-S[X,\hat G;g_{\mu\nu}]},
\label{35}
\end{equation}
which contains a path integral of the $N=2$ nonlinear sigma model
\begin{equation}
Z(\beta,\phi)=\int_{T^1} {\cal D} X\,\E^{-S[X,\hat G;g_{\mu\nu}]}.
\label{36}
\end{equation}
As explained in the previous section, these path integrals are
completely under control, and thus one can proceed with concrete
applications. An interesting feature of this worldline approach to
vector ($p=1$) and general antisymmetric tensor fields is that on
top of the proper time $\beta$ there is a new modular parameter
$\phi$. It is related to the Wilson loop variable by
$$
w=\exp\biggl(\I\int_0^1a\D\tau\biggr)=\E^{\I\phi}\,.
$$
The integration $\phi\in[0,2\pi]$ has the effect of projecting
onto the correct physical sector described by the $(p+1)$-form
$F_{p+1}$. It is interesting to note that the parameter $\phi$ can
be eliminated from the action by a field redefinition of the
fermions, which then acquire different boundary conditions:
$\psi^a(1)=-\E^{\I\phi}\psi^a(0)$. Averaging over $\phi$ can then
be interpreted as averaging over all possible boundary conditions
of the fermions. Note that at $\phi=\pi$ a zero mode of the free
fermionic kinetic term appears: at this point the fermions have
periodic boundary conditions and constant fields $\psi^a_0$ are
zero modes.

The worldline representation  of the effective action in
(\ref{35}) is quite explicit, and can be used to compute
$\Gamma^{QFT}_{p}$ in some approximation (the exact evaluation
with an arbitrary background metric is impossible to achieve with
current techniques). For example, the perturbative evaluation of
the path integral for the $N=2$ nonlinear sigma model in
(\ref{36}) at order $\beta^2$ can be carried out without any
ambiguity, as already explained in section 2. It allows to
identify the first three Seeley--DeWitt coefficients $a_0$, $a_1$,
$a_2$ for an arbitrary $p$-form in arbitrary dimensions. They
appear as follows
$$
\Gamma^{\mathrm{QFT}}_{p}[g_{\mu\nu}]=
-\frac12\int_0^\infty\frac{\D\beta}{\beta}
\int\frac{\D^Dx\sqrt{g(x)}}{(2\pi\beta)^{D/}}\,
\bigl(a_0(x)+a_1(x)\beta+a_2(x)\beta^2+\ldots\bigr)
$$
and their values have been reported in \cite{Ba:2005vk}. The cases
for $p=0,1,2$ were already known in the literature
\cite{DW,DeWitt:2003pm}, but the cases for $p\geq3$ are new. These
coefficients can in principle be obtained by specializing the
known Seeley--DeWitt coefficients for a scalar field coupled to an
arbitrary connection to the case under study, and performing the
necessary index contractions. However, the latter task is quite
laborious. The worldline representation maps this problem into the
problem of computing some worldline fermion correlators, and makes
the computation quite easy and efficient.

A technical point worth of commenting upon is related to the
appearance of a singularity on the U(1) moduli space. This
singularity appears precisely at the point $\phi=\pi$, where
perturbative zero modes arise for the fermions. It is convenient
for the present discussion to use the Wilson loop variable $w$  in
place of $\phi$, and switch to an operatorial picture. Then the
effective action in (\ref{35}) can be rewritten as follows
\begin{equation}
\Gamma^{\mathrm{QFT}}_{p}[g_{\mu\nu}]=
-\frac12\int_0^\infty\frac{\D\beta}{\beta} \oint_{\gamma}\frac{\D
w}{2\pi\I w}\,\frac{w}{(1+w)^2}\, \Tr\left[w^{\hat
N-(p+1)}\E^{-\beta hat H}\right],
\label{lastea}
\end{equation}
where $\hat N$ is the (anti)fermion number operator
$\hat\psi^a\hat\psi_a^\dagger$, and the integration region of the
Wilson loop variable $w$ is the unit circle $\gamma$ in the
complex $w$-plane. The singular point $\phi=\pi$ is now mapped to
$w=-1$. In particular, the presence of the susy ghost determinant
$\dfrac{w}{(1+w)^2}$ makes this pole rather dangerous. The
prescription devised in \cite{Ba:2005vk} is to deform the contour
to exclude the point $w=-1$, and use contour integration to
evaluate the integrals, see Fig. 4.

\begin{figure}[t]
\begin{center}
\scalebox{.5}{
\includegraphics*[1pt,1pt][288pt,288pt]{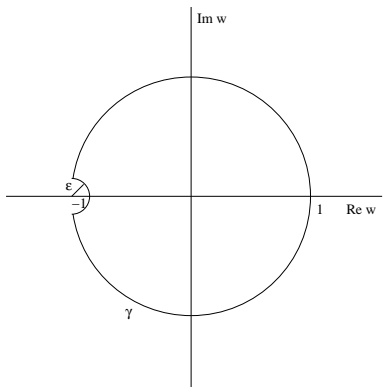}}
\end{center}
\vspace{-8mm} \caption{Regulated contour on the U(1) moduli
space.}
\end{figure}

This prescription permits the calculation of the Seeley--DeWitt
coefficients $a_0$, $a_1$  and $a_2$ for all $p$-forms in
arbitrary dimensions, reproducing in particular the known results
for the cases of $p=0,1,2$. Moreover, it sheds an interesting
light on the issue of quantum equivalence of dual $p$-forms
\cite{Duff:1980qv,Siegel:1980ax}. A massless $p$-form is expected
to be equivalent to a massless $(D-p-2)$-form. Replacing $p$  with
$(D-p-2)$ in (\ref{lastea}) gives directly the effective action
for a massless $(D-p-2)$-form. This replacement only produces a
change $q\to -q$. This change that can be undone by a subsequent
change of integration variables $w\to w'=1/w$. This would seem to
prove the exact equivalence.  However, the new integration
variable has a modified contour of integration which includes the
pole, see Fig. 5, so that the total mismatch between the effective
actions of dual differential forms is related to the residue at
the pole $w=-1$.

\begin{figure}[t]
\begin{center}
\scalebox{.5}{
\includegraphics*[1pt,1pt][288pt,288pt]{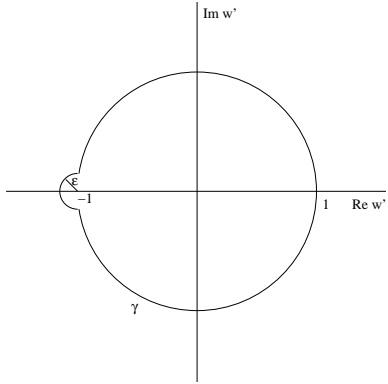}}
\end{center}
\vspace{-8mm} \caption{Contour for the dual differential form.}
\end{figure}

In even dimension the mismatch is a local term, proportional to
the Euler topological density, which affects that particular
Seeley--DeWitt coefficient with the same dimensions of the Euler
term. This mismatch was first noticed in four dimensions in
\cite{Duff:1980qv}. It contributes to the local terms that are
usually subtracted when one renormalizes the effective action, and
thus, according to \cite{Siegel:1980ax}, it does not really
destroy duality. In odd dimension the mismatch is also topological
and corresponds to the so-called Ray--Singer torsion, as
discovered in \cite{Schwarz:1984wk}. This mismatch can be
interpreted as the additional contribution of a $(D-1)$-form gauge
potential, which however carries no degrees of freedom. Exact
formulas can be found in \cite{Ba:2005vk}.

In this section we have described an application of the worldline
approach to arbitrary antisymmetric tensor fields coupled to
gravity. This approach can of course be used to compute some
one-loop amplitudes with a certain efficiency as well, see
\cite{Ba:2005vk}. The particular case $p=1$ describes a photon
coupled to gravity. Previous worldline descriptions of spin 1
particles in $D=4$ dimensions have been considered in
\cite{Strassler:1992zr} and \cite{Reuter:1996zm}. In those
references only a rigid $N=2$ linear sigma model was used,
together with a limiting procedure necessary to achieve the
propagation of the correct degrees of freedom. This limiting
procedure is not particularly elegant, but it allows the inclusion
of Yang--Mills backgrounds. It is not clear how to include the
latter using the $N=2$ spinning particle with local supersymmetry
described above.

\bigskip
\noindent{\small I would like to thank all the collaborators that
at various stages joined me in the study and in the applications
of the path integral in curved space: Paolo Benincasa, Olindo
Corradini, Hari Dass, Simone Giombi, Koenraad Schalm, Christian
Schubert, Peter van Nieuwenhuizen, Andrea Zirotti.}

\end {document}